\RequirePackage{lineno}
\documentclass[preprint,12pt]{elsarticle}

\usepackage{amssymb}
\usepackage{graphicx}
\usepackage{color}

\newcommand{\jpsi}{\rm J/$\psi$}

\newcommand{\slfrac}[2]{\left.#1\right/#2}

\journal{Physics Letters B}

\begin{document}
\begin{frontmatter}



\title{J/$\psi$ production in proton-nucleus collisions\hfil\break 
at 158 and 400 GeV}


\author[1]{R.~Arnaldi}
\author[2,3]{K.~Banicz} 
\author[4]{J.~Castor} 
\author[5]{B.~Chaurand}
\author[6]{W.~Chen} 
\author[7]{C.~Cical\`o}  
\author[8]{A.~Colla} 
\author[1]{P.~Cortese}
\author[2,3]{S.~Damjanovic}
\author[2,9]{A.~David} 
\author[10]{A.~de~Falco} 
\author[4]{A.~Devaux }
\author[11]{L.~Ducroux} 
\author[12]{H.~En'yo}
\author[4]{J.~Fargeix}
\author[8]{A.~Ferretti} 
\author[10]{M.~Floris} 
\author[2]{A.~F\"orster}
\author[4]{P.~Force}
\author[2,4]{N.~Guettet}
\author[11]{A.~Guichard} 
\author[13]{H.~Gulkanian} 
\author[12]{J.~M.~Heuser}
\author[2,9]{M.~Keil} 
\author[6]{Z.~Li}
\author[2]{C.~Louren\c{c}o}
\author[9]{J.~Lozano} 
\author[4]{F.~Manso} 
\author[2,9]{P.~Martins}  
\author[7]{A.~Masoni}
\author[9]{A.~Neves}
\author[12]{H.~Ohnishi} 
\author[1]{C.~Oppedisano}
\author[9]{P.~Parracho} 
\author[11]{P.~Pillot} 
\author[13]{T.~Poghosyan}
\author[10]{G.~Puddu} 
\author[2]{E.~Radermacher}
\author[2,9]{P.~Ramalhete} 
\author[2]{P.~Rosinsky} 
\author[1]{E.~Scomparin}
\author[9]{J.~Seixas} 
\author[10]{S.~Serci} 
\author[2,9]{R.~Shahoyan} 
\author[9]{P.~Sonderegger}
\author[3]{H.~J.~Specht} 
\author[11]{R.~Tieulent} 
\author[10]{A.~Uras} 
\author[10]{G.~Usai} 
\author[9]{R.~Veenhof}
\author[9]{H.~K.~W\"ohri}

\address[1]{INFN, Sezione di Torino, Italy}
\address[2]{CERN, 1211 Geneva 23, Switzerland}
\address[3]{Physikalisches~Institut~der~Universit\"{a}t Heidelberg,~Germany}
\address[4]{Universit\'e Blaise Pascal and CNRS-IN2P3, Clermont-Ferrand, France}
\address[5]{LLR, Ecole Polytechnique and CNRS-IN2P3, Palaiseau, France}
\address[6]{BNL, Upton, NY, USA}
\address[7]{INFN, Sezione di Cagliari, Italy}
\address[8]{Dipartimento di Fisica Sperimentale dell' Universit\`a di Torino and INFN, Torino,~Italy}
\address[9]{Instituto Superior T\'ecnico, Lisbon, Portugal}
\address[10]{Dipartimento di Fisica dell' Universit\`a di Cagliari and INFN, Cagliari, Italy}
\address[11]{IPNL, Universit\'e Claude Bernard Lyon-I and CNRS-IN2P3, Villeurbanne, France}
\address[12]{RIKEN, Wako, Saitama, Japan}
\address[13]{YerPhI, Yerevan Physics Institute, Yerevan, Armenia}

\begin{abstract}
The NA60 experiment has studied \jpsi\ production in \mbox{p-A} collisions at
158 and 400 GeV, at the CERN SPS. Nuclear effects on the \jpsi\ yield have 
been estimated from the A-dependence of the production cross section ratios 
$\sigma_{\rm J/\psi}^{A}/\sigma_{\rm J/\psi}^{Be}$ (A=Al, Cu, In, W, Pb, U).
We observe a significant nuclear suppression of the \jpsi\ yield per \mbox{nucleon-nucleon}
collision, with a larger effect  at lower incident energy, and we compare this
result with previous observations by other fixed-target experiments. 
An attempt to disentangle the different contributions to the observed suppression has been carried out by studying the dependence of nuclear effects on $x_2$, the fraction of the nucleon momentum carried
by the interacting parton in the target nucleus.
\end{abstract}


\end{frontmatter}


\section{Introduction}
\label{sec:intro}

The study of charmonium production in hadronic collisions is an interesting 
test of our understanding of the physics of strong interactions. While the production 
of the $c\overline c$ pair can be addressed in a perturbative-QCD approach, 
the subsequent binding of the pair is an essentially non-perturbative process, 
involving soft partons and occurring on a rather long timescale ($>1$ fm/c).
Various theoretical approaches have been proposed
(see~\cite{Lan09} for a recent review), but a
satisfactory description of \jpsi\ production  in \mbox{p-p} collisions is still 
missing~\cite{QWG11}.
In \mbox{p-A} interactions, the heavy-quark pair is created in the nuclear medium, and
the study of its evolution towards a bound state can add significant constraints to the models.
For example, the strength of the interaction between the evolving $c\overline c$ pair 
and the target nucleons, that can lead to a break-up of the pair and consequently to a 
suppression of the \jpsi\ yield, may depend on its quantum state at the production level
(color-octet or color-singlet), and on the kinematic variables of the pair~\cite{Vog02,Kop91}.
In addition to final-state effects, also initial-state effects may influence the observed \jpsi\ yield in \mbox{p-A}. 
In particular, parton shadowing in the target nucleus~\cite{Esk09} may suppress (or enhance, in the case 
of anti-shadowing) the probability of producing a \jpsi, while the energy loss of the incident parton~\cite{Gav92} in the
nuclear medium, prior to $c\overline c$ production, may
significantly alter the \jpsi\ cross section and kinematic distributions. Furthermore,
effects as final-state energy loss and the presence of an intrinsic charm component in the proton
may also play a significant role~\cite{Vog00}.
Clearly, the correct understanding and the disentangling of the various nucleus-related 
effects on \jpsi\ production is a non-trivial task, which poses significant challenges to
theory, but at the same time offers important insights on the \jpsi\ production and 
interaction mechanisms.
Finally,
a suppression of the \jpsi\ has been proposed a long time ago as a signature
of the formation, in ultrarelativistic nucleus-nucleus collisions, of a state where 
quarks and gluons are deconfined (Quark-Gluon Plasma)~\cite{Sat86}. 
Results from \mbox{p-A} collisions, taken in the same kinematic conditions of 
\mbox{A-A}, and properly extrapolated to nucleus-nucleus collisions, 
are therefore necessary to calibrate the contribution of the various cold nuclear 
matter effects to the overall observed suppression~\cite{Ram05,Arn07}. 

Having to deal with a rather complicated interplay of various physical
processes, the availability of accurate sets of data, spanning large intervals in
the incident proton energy, and covering large $x_{\rm F}$ and $p_{T}$ regions, is 
essential for a thorough understanding of the involved mechanisms. At fixed target
energies, high-statistics \jpsi\ samples have been collected in recent years by the DESY 
experiment HERA-B~\cite{Abt09}, at 920 GeV incident energy, by E866~\cite{Lei00} at FNAL 
at 800 GeV, and by the CERN-SPS experiment NA50 at 400 and 450 GeV~\cite{Gon06}. 

In this Letter, we present an extension of these measurements towards lower energies, carried out by the NA60 
experiment~\cite{Arn09}.
\jpsi\ production has been studied in \mbox{p-A} collisions at 158 GeV, the same energy used in \mbox{A-A}
collisions at the CERN SPS. Data have also been taken with a 400 GeV beam,
in order to provide a result that can be compared with previous sets of \mbox{p-A} data taken 
by the NA50 experiment~\cite{Gon06}. We have performed a systematic study of nuclear effects by analyzing the
$A-$dependence of the \jpsi\ production cross section on seven different target nuclei (Be, Al, Cu, In, W, Pb and U).
The results are relative to a region close to midrapidity, and are presented differentially in $x_F$ and $x_2$ and compared with 
previous results from the higher-$\sqrt{s}$ measurements mentioned above. The influence of parton shadowing in the 
nuclear targets on the results is also discussed, based on various recent parametrizations of this 
effect~\cite{Esk09,Hir04,Esk99,Esk08}.  
A comparison between our results at the two energies as a function of $x_2$, the fraction of the nucleon momentum 
carried by the parton in the nuclear target that produces the \jpsi, is carried out. By doing so, the shadowing effects can be factorized and the importance of the other nuclear effects can be investigated.
 
\section{Data analysis}
\label{sec:data}

The NA60 experiment has measured muon pair production in \mbox{p-A} and \mbox{A-A} collisions at the CERN SPS.
Its experimental apparatus was based on a muon spectrometer (MS), positioned downstream
of a hadron absorber with a total thickness of 12 nuclear interaction lengths ($\lambda_I$). The MS is coupled to a vertex spectrometer (VT) based on Si pixel detectors. The experiment triggered on muon pairs detected in the MS, which were then matched, during reconstruction, to the corresponding tracks in the VT.
For details on the detector set-up and matching between the MS and VT we refer to~\cite{Arn09}.

The \jpsi\ mesons have been identified through their decay
to a muon pair.
In the \mbox{p-A} data taking, the SPS proton beam, with an average
intensity of 5$\cdot$10$^8$ s$^{-1}$, was hitting a target system composed of nine sub-targets 
with thicknesses between 0.005 and 0.012 $\lambda_I$, 
with relative spacing between targets from 0.8 to 1 cm. 
To limit the possible influence of position-dependent systematic effects on the 
measurement of the nuclear dependence of the \jpsi\ yield, the subtargets were placed in a
mixed-A order, namely  Al, U, W, Cu, In, Be, Be, Be and Pb.

The analysis described in this Letter has been carried out on a data sample consisting of 
2.8$\cdot$10$^6$ events at 158 GeV, and 1.5$\cdot$10$^6$ events at 400 GeV, containing  
a dimuon reconstructed in the MS. 
The mass region m$_{\mu\mu}>$2.85 GeV/c$^2$, dominated by \jpsi\ 
decays, contains 3.2$\cdot$10$^4$ and 2.1$\cdot$10$^4$ events at 158 and 400 GeV, respectively.
For about 50\% of these events, each of the two MS tracks can be matched in direction and momentum with a track in the VT. Such a value of the dimuon matching efficiency is due the smaller solid angle coverage of the VT with respect to the MS and to the inefficiency of the pixel sensors as will be discussed below. A Monte-Carlo simulation has shown that 
the contamination from events where at least one of the MS tracks is matched to a wrong VT track is negligible
in the \jpsi\ mass region.

\begin{figure}[htbp]
\centering
\resizebox{0.8\textwidth}{!}
{\includegraphics*[bb=0 0 539 383]{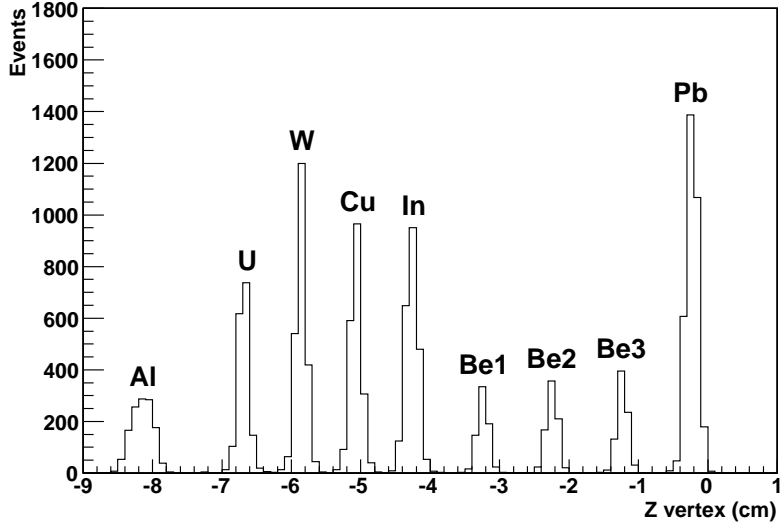}}
\caption{The distribution of the longitudinal coordinate of the point of closest approach of the opposite sign muon pair, 
for m$_{\mu\mu}>$2.85 GeV/c$^2$.}
\label{fig:0}
\end{figure}

For events containing a pair of hard muons from the decay of a \jpsi\  (the average muon momentum is 27 GeV/c), 
the point of closest approach of the two muons gives an accurate estimate of their production point, 
with a resolution of $\sim$ 650 $\mu$m. In Fig.~\ref{fig:0} we show the distribution of the longitudinal coordinates of the dimuon 
production vertex, for m$_{\mu\mu}>$2.85 GeV/c$^2$. The peaks corresponding to the position of the various production targets are clearly
visible and well separated, showing that  the nuclear target where the 
\jpsi\ has been produced is unambiguously determined.

The MS detects \jpsi\ produced in the rapidity range 3$<y_{\rm lab}<$4. However, the acceptance of the VT
for matched tracks from the \jpsi\ decay is target-dependent, the more downstream targets covering smaller rapidities. 
We have therefore selected events produced in the range 3.2$<y_{\rm lab}<$3.7, which is covered for all the targets. 
The cut -0.5$<\cos\theta_{\rm CS}<$0.5, on the polar angle of the decay muons in the Collins-Soper reference frame 
has also been applied, in order to remove events at the edge of
the MS's acceptance. The transverse momentum coverage for the \jpsi\ measurement extends down to zero $p_{\rm T}$.
The acceptance varies by about 30\% in the $p_{\rm T}$ range accessible with the collected statistics 
($p_{\rm T}\lesssim$ 3 GeV/c).

\begin{figure}[htbp] 
\centering
\resizebox{0.8\textwidth}{!}
{\includegraphics*[bb=0 0 520 350]{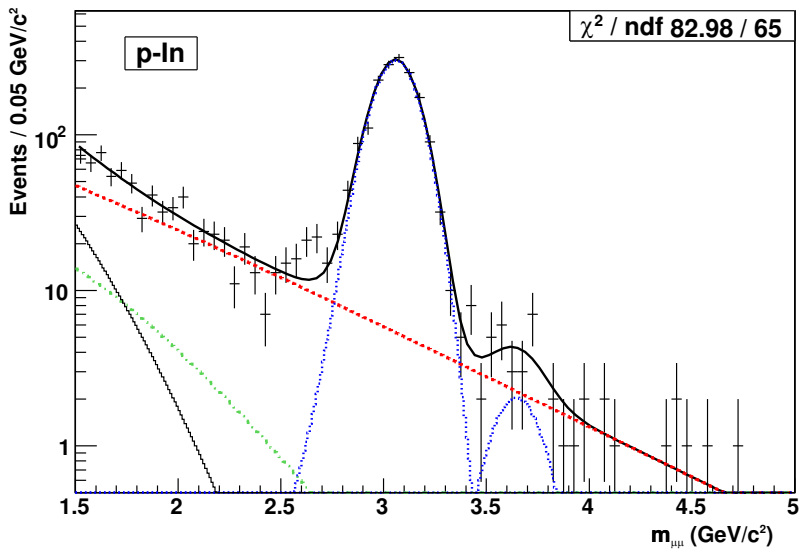}}
\caption{Fit to the $\mu^+\mu^-$ invariant mass spectrum for \mbox{p-In} collisions
at 158 GeV. The dashed line represents the Drell-Yan process, the dotted lines the charmonium resonances, the dashed-dotted line the 
$D\overline{D}$ contribution, the thin continuous line the combinatorial background. The thick continuous line is the sum of all the 
contributions.}
\label{fig:1}
\end{figure}

The \jpsi\ signal has been extracted for each target nucleus at the two energies by fitting the opposite sign mass spectra, in the region 
m$_{\mu\mu}>$1.5 GeV/c$^2$, with a superposition of the mass shapes of the expected dimuon sources. 
These include the Drell-Yan process (DY),
the semi-leptonic decays of correlated D-meson pairs ($D\overline D$), the \jpsi\ and the $\psi'$ resonances. 
The expected mass distributions for the DY and $D\overline D$ contributions have been
calculated with PYTHIA~\cite{Sjo01}, using the GRV94LO~\cite{Glu95} parton distribution functions.
The \jpsi\ events have been generated, for the
400 GeV data, using the $y$ and $p_{\rm T}$ distributions  measured with good accuracy by NA50~\cite{Ale04}. 
At 158 GeV, the differential distributions have been tuned directly on the data, using an iterative procedure.
The events have been tracked through the set-up and then
reconstructed with the same algorithm used for real data. The contribution from the 
combinatorial pair background due to $\pi$ and $K$ decays (completely negligible in the \jpsi\ mass region) has been estimated through an 
event mixing technique~\cite{Arn09}. 
In Fig.~\ref{fig:1} we show, as an example, the opposite-sign dimuon invariant mass distribution, relative to the kinematical domain 
specified above, for \mbox{p-In} interactions at 158 GeV. 
The quality of all the fits to the invariant mass spectra
is satisfactory (with $\chi^2/ndf$ ranging from 0.7 to 1.5). The mass resolution at the \jpsi\ peak is $\sim$70 MeV/c$^2$, 
and the number of \jpsi\ events ranges from $\sim$800 to $\sim$2000, depending on the target. The number of continuum events
under the \jpsi\ peak is very small ($<$4\%), and is dominated by the DY contribution.

Nuclear
effects have been parametrized by fitting the $A-$dependence of the production cross section
with the simple power law $\sigma_{\rm J/\psi}^{pA} = \sigma_{\rm J/\psi}^{pp}\cdot
A^{\alpha}$, and then studying the evolution of $\alpha$ as a function of various kinematic variables. 
Alternatively, nuclear effects have been quantified by fitting the data in the framework 
of the Glauber model, 
having as input parameters 
the density distributions for the various nuclei. 
The model gives as 
output the so-called \jpsi\ absorption cross section $\sigma_{\rm J/\psi}^{abs}$. 
Clearly, both $\alpha$ and $\sigma_{\rm J/\psi}^{abs}$ represent effective quantities, 
including the contribution of the various sources of nuclear effects detailed in the Introduction.

The nuclear effects on \jpsi\ production have been evaluated starting from the cross section measured for each target, 
normalized to the cross section for the lightest one (Be):
\begin{equation}
\frac{\slfrac{\sigma_{A}^{\rm J/\psi}}{A}}{\slfrac{\sigma_{Be}^{\rm J/\psi}}{A_{Be}}}= 
\slfrac {\frac{\slfrac{N_{A}^{\rm J/\psi}}{A}}{N_{A}^{\rm inc}\cdot N_{A}^{\rm targ}\cdot
\mathcal{A}_{A}\cdot\epsilon_{A}}} {\frac{\slfrac{N_{Be}^{\rm J/\psi}}{A_{Be}}}{N_{Be}^{\rm inc}\cdot N_{Be}^{\rm targ}\cdot
\mathcal{A}_{Be}\cdot\epsilon_{Be}}}
\end{equation}

\noindent where, for the target with mass number $A$, $N_{A}^{\rm J/\psi}$ is the number of \jpsi\ events, $N_{A}^{\rm inc}$ is the number of
incident protons, $N_{A}^{\rm targ}$ is the number of target nuclei per unit surface, $\mathcal{A}_{A}$ is the \jpsi\ acceptance and 
$\epsilon_{A}$ is the detection efficiency. 
When building these ratios, the results obtained with the three Be targets have been averaged.
The use of relative cross sections brings several advantages. 
In particular, the number of incident protons cancels out (a small attenuation factor, due to the inelastic 
scattering of beam particles, which is 6\% for the most downstream target, has been corrected for) since all the targets were simultaneously exposed to the beam, which had a transverse dimension much smaller than that of the targets. 
Furthermore, 
the fraction of the detection efficiency related to the MS also cancels out. In fact, this detector cannot
distinguish the target where the dimuon has been produced. This is due to the presence of the thick hadron absorber, 
to the large distance between the target system and the MS tracking chambers (6 to 16 m), and to the closely spaced targets. 
Therefore, the muon detection efficiency is independent of the production target. 


Contrary to the MS case, the angular acceptance covered by the VT is slightly target dependent. Therefore, non-uniformities in the efficiency of a certain pixel plane lead to a rapidity-dependent inefficiency which is different from target to target. The efficiency of each of the 15 VT planes has been estimated using a modified track reconstruction algorithm that 
excludes the plane under study. For each reconstructed track we then check the presence of a hit in that plane in a 
fiducial region around the intersection of the track with the plane. Efficiency values are calculated on a run-per-run 
basis ($\lesssim$2 hours), for sub-regions of the pixel plane down to a size of $\sim$0.2 mm$^2$, if the track statistics in the sub-region 
under study is large enough (more than 40 tracks). Otherwise, contiguous regions are grouped in order to reach a 
statistically significant track sample. The distribution of the efficiency values for the various regions of the VT is peaked 
at $\sim$90\%, with 82\% of the 
detector area having an efficiency larger than 60\%.
Finally, the \jpsi\ acceptances, estimated by means of the Monte-Carlo simulation,  
range between 14.9\% and 23.0\% 
in the kinematic range 3.2$<y_{lab}<$3.7, $|\cos\theta_{\rm CS}|<$0.5. The quoted rapidity range 
corresponds to the center-of-mass rapidity windows 0.28$<y<$0.78 at 158 GeV and $-0.17<y<$0.33 at 400 GeV.

In Table~1 we summarize, target by target and for the two energies under study, the relative values of the \jpsi\ 
acceptance times detection efficiencies. For completeness, we also report the relative integrated luminosities
$\mathcal{L}_{A}/\mathcal{L}_{Be}= (N_{A}^{\rm inc}\cdot N_{A}^{\rm targ})/(N_{Be}^{\rm inc}\cdot N_{Be}^{\rm targ})$.

\begin{table}
\begin{center}
\begin{tabular}{|c|c|c|c|}
\hline 
 & $\mathcal{L}_{A}/\mathcal{L}_{Be}$  & $(\mathcal{A}_{A}\cdot\epsilon_{A})/(\mathcal{A}_{Be}\cdot\epsilon_{Be})(158)$ & 
$(\mathcal{A}_{A}\cdot\epsilon_{A})/(\mathcal{A}_{Be}\cdot\epsilon_{Be})(400)$  \tabularnewline
\hline
\hline 
Al & 1.048$\pm$0.002 & 0.58$\pm$0.02 & \tabularnewline
\hline 
U & 0.184$\pm$0.003 & 0.73$\pm$0.02 & 0.81$\pm$0.02\tabularnewline
\hline 
W & 0.261$\pm$0.005 & 0.80$\pm$0.03 & 0.89$\pm$0.03\tabularnewline
\hline 
Cu & 0.529$\pm$0.003 & 0.89$\pm$0.02 & 0.93$\pm$0.02\tabularnewline
\hline 
In & 0.320$\pm$0.002 & 0.95$\pm$0.03 & 0.97$\pm$0.03\tabularnewline
\hline 
Be & 1.000$\pm$0.005 & 1.00$\pm$0.02 & 1.00$\pm$0.02\tabularnewline
\hline 
Be & 0.991$\pm$0.005 & 1.05$\pm$0.02 & 1.01$\pm$0.02\tabularnewline
\hline 
Be & 0.991$\pm$0.005 & 1.03$\pm$0.02 & 1.00$\pm$0.02\tabularnewline
\hline 
Pb & 0.269$\pm$0.002 & 0.96$\pm$0.04 & 0.95$\pm$0.05\tabularnewline
\hline
\end{tabular}
\label{tab:1}
\end{center}
\caption{The relative values, target by target, of the integrated luminosities, and of the products $\mathcal{A}_{A}\cdot\epsilon_{A}$. We have chosen as reference value the most upstream Be target. The Al target was not in place during the data taking at 400 GeV.}
\end{table}

\section{Results}
\label{sec:resu}

\begin{figure}[htbp]
\centering
\resizebox{0.8\textwidth}{!}
{\includegraphics*[bb=0 0 539 383]{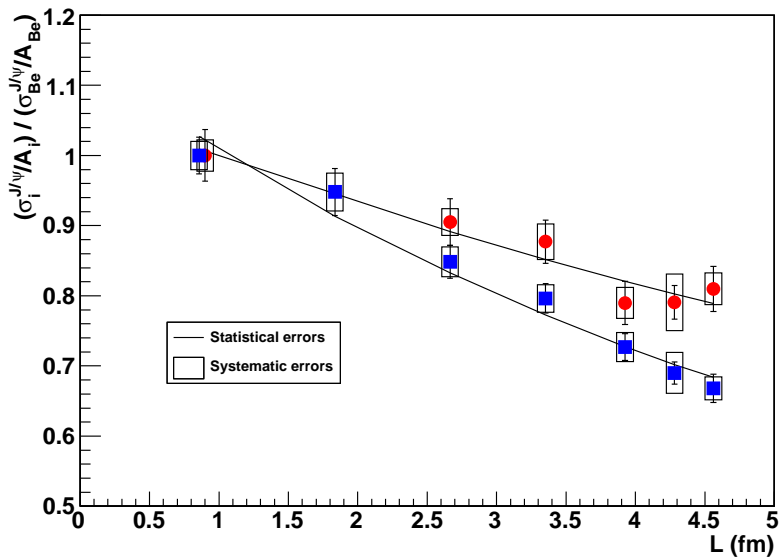}}
\caption{Cross sections for \jpsi\ production in \mbox{p-A} collisions, normalized to the \mbox{p-Be} \jpsi\ cross section. The squares
represent the 158 GeV results, the circles refer to 400 GeV. The lines are the fit results.}
\label{fig:2}
\end{figure}

In Fig.~\ref{fig:2} we present the cross section ratios $(\sigma_{i}^{\rm J/\psi}/A_{i})/(\sigma_{Be}^{\rm J/\psi}/A_{Be})$, 
at 158 and 400 GeV, where $A_i$ is the nuclear mass number of target $i$. The
results are shown as a function of $L$, the mean thickness of nuclear matter crossed by the \jpsi\ in its way through the 
target nucleus. The $L$ values have been computed with the Glauber model, using realistic density distributions for the various 
nuclei~\cite{DeV87}.
The quoted systematic uncertainties include contributions, quadratically combined, coming from the uncertainty on
i) the measurement of the target thicknesses ($\le$1.5\%), ii) the \jpsi\ acceptance, due to the choice of the rapidity
distribution adopted in the Monte-Carlo calculation ($\le$1.5\%), iii) the efficiency calculation ($\le$3.0\%). 
This last, and most important, contribution has been obtained varying by $\pm$10\% the estimated efficiency values of the VT pixel detectors. 
We only quote, for each incident energy, the fraction of
the systematic uncertainty which is not common to all the points, the only one relevant when plotting relative cross
sections. Fig.~\ref{fig:2} shows, for both datasets, a suppression of the \jpsi\ yield when moving from light to 
heavy targets, and, in
particular, a larger suppression for the 158 GeV data sample. Using the Glauber model, we have estimated 
$\sigma_{\rm J/\psi}^{abs}(158\, \rm{GeV})= 
7.6\pm 0.7 \rm {(stat)} \pm 0.6 \rm{(syst)}$ mb and $\sigma_{\rm J/\psi}^{abs}(400\, \rm{GeV})= 
4.3\pm 0.8 \rm {(stat)} \pm 0.6 \rm{(syst)}$ mb. With the power-law parameterization we get 
$\alpha_{\rm J/\psi}(158\, \rm{GeV})=0.882 \pm 0.009 \rm {(stat)} \pm 0.008 \rm{(syst)}$ and
$\alpha_{\rm J/\psi}(400\, \rm{GeV})=0.927 \pm 0.013 \rm {(stat)} \pm 0.009 \rm{(syst)}$. We note that the 400 GeV values are in good agreement with
results from NA50 at the same energy ($\sigma_{\rm J/\psi}^{abs}= 4.6\pm 0.6$ mb~\cite{Gon06}), in a similar rapidity range. We also
stress that the reported values have not been corrected for shadowing effects and therefore represent a global effective estimate of the
nuclear effects on the \jpsi\ yield.

\begin{figure}[htbp]
\centering
\resizebox{0.8\textwidth}{!}
{\includegraphics*[bb=0 0 539 383]{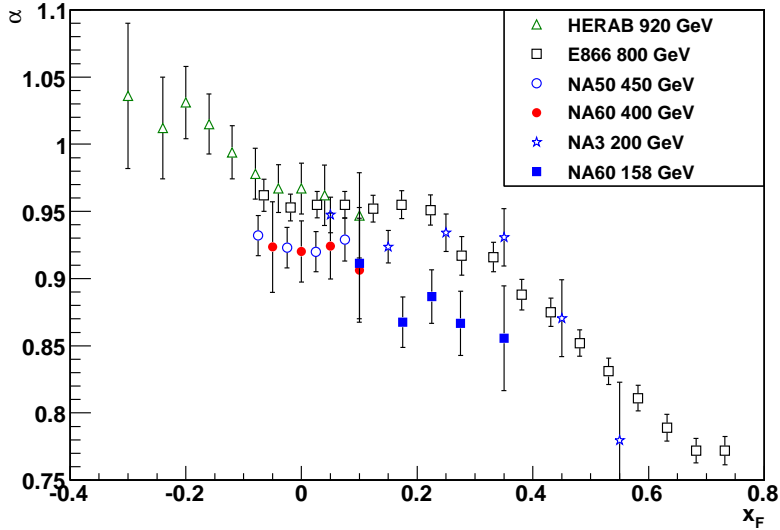}}
\caption{The $x_{\rm F}$ dependence of the $\alpha$ parameter. Open triangles correspond to HERA-B results, 
open squares to E866, open circles to NA50 (450 GeV), closed circles to NA60 (400 GeV), closed squares to NA60 (158 GeV), 
open stars to NA3. The error bars represent the quadratic sum of statistical and systematic uncertainties.}
\label{fig:3}
\end{figure}

To get further insight on the dependence of nuclear effects on the \jpsi\ kinematic variables we compare in
Fig.~\ref{fig:3}, as a function of $x_{\rm F}$, the $\alpha$ values obtained in this analysis with those from previous experiments in the
fixed-target energy range. For this purpose, the NA60 sample has been subdivided into 5 $x_{\rm F}$ bins at 158 GeV 
(4 at 400 GeV), covering the region $0.05<x_{\rm F}<0.40$ ($-0.075<x_{\rm F}<0.125$). In Table~2 we present the
corresponding $\alpha$ values.

\begin{table}
\begin{center}
\begin{tabular}{|c|c|c|}
\hline
Energy (GeV) & $x_{F}$ & $\alpha$\tabularnewline
\hline
\hline
158 & $0.05\div0.15$ & $0.911\pm0.019\pm0.036$\tabularnewline
\hline
158 & $0.15\div0.20$ & $0.868\pm0.016\pm0.009$\tabularnewline
\hline
158 & $0.20\div0.25$ & $0.887\pm0.017\pm0.008$\tabularnewline
\hline
158 & $0.25\div0.30$ & $0.867\pm0.019\pm0.010$\tabularnewline
\hline
158 & $0.30\div0.40$ & $0.856\pm0.025\pm0.027$\tabularnewline
\hline
400 & $-0.075\div-0.025$ & $0.924\pm0.026\pm0.021$\tabularnewline
\hline
400 & $-0.025\div0.025$ & $0.920\pm0.019\pm0.012$\tabularnewline
\hline
400 & $0.025\div0.075$ & $0.924\pm0.020\pm0.013$\tabularnewline
\hline
400 & $0.075\div0.125$ & $0.906\pm0.028\pm0.025$\tabularnewline
\hline
\end{tabular}
\label{tab:2}
\end{center}
\caption{$\alpha$ as a function of $x_{\rm F}$ for the 158 and 400 GeV data sample. The first quoted uncertainty
is statistical, the second is the systematic one.}
\end{table}

Two main features emerge from this comparison. First, when going from negative towards positive $x_{\rm F}$, $\alpha$ steadily decreases. 
This effect was already known from the data of HERA-B and E866, taken at rather similar incident 
proton energies (920, 800 GeV), and a similar effect might be present in our results at 158 GeV, even if the size of the errors is not negligible. 
Second, at a constant $x_{\rm F}$, the $\alpha$ values are lower when the incident proton energy is smaller, as can be seen when comparing the HERA-B/E866 results with our results at 400 and 158 GeV.  
On the other hand, it is also worth mentioning that from NA3 results on \jpsi\ 
production at 200 GeV~\cite{Bad82} one extracts $\alpha$ values which are in partial contradiction with these observations,
being similar to those obtained with the higher energy data samples (HERA-B/E866).
We also note that in our calculation of the NA3 $\alpha$ values, performed starting from their measured \jpsi\ cross section ratios between
\mbox{p-Pt} and \mbox{p-p} collisions, the small bias~\cite{Sha01} induced by the use of a light target in the determination of 
$\alpha$ has been corrected for. 

A satisfactory theoretical interpretation of the complex observed pattern is missing for the
moment. Various works have underlined the importance of several effects, including final state break-up, parton shadowing, initial and final
state energy loss, and the presence of a charm component in the nucleon wavefunction~\cite{Vog00,Arl09, Lou09}. However, 
the relative weight of these effects is still under debate.

The new results from NA60 at 400 and 158 GeV have been obtained with the same experimental apparatus and in very similar running 
conditions. Their direct comparison may therefore be a clean testing ground for models. Interesting information can also be 
obtained by a simple comparison of nuclear effects as a function of various kinematic variables. 
As an example, we now consider the $x_2$ dependence of the $\alpha$ parameter, $x_2 = m_T/\sqrt{s}\cdot \exp(-y)$ being the fraction of the
momentum of the target nucleon carried by the parton which produces the \jpsi\footnote{Such a relation, commonly adopted, implies that the 2$\rightarrow$2 kinematics of the main J/$\psi$ production process $gg\rightarrow {\rm J}/\psi g$ is, effectively, that of a 2$\rightarrow$1 process ($gg\rightarrow {\rm J}/\psi$) i.e. the final state gluon is very soft. The alternative 2$\rightarrow$2 approach is discussed in~\cite{Fer09}. \\
We also note that the relation strictly holds when $p_{T}\ll p_{L}$ in the c.m. frame. Due to the low $\sqrt{s}$, this is only approximately true and induces a smearing in the estimation of $x_{2}$. However, in the investigated kinematic domain, the effect is similar for the two energies. The resulting $\sim 10\%$ shift in the $x_2$ values does not affect the discussion in the text.}. 
This kinematic variable is particularly interesting since 
$x_2$ is the quantity that determines the amount of shadowing in the target nucleus 
(or anti-shadowing in our kinematic range).
Furthermore, the center of mass energy of the \jpsi-N system, which is a relevant quantity for the final state break-up of the \jpsi\ in the cold nuclear medium, is, to a very good approximation,
a function of $x_2$ alone ($\sqrt{s_{\rm J/\psi N}}\sim \slfrac{m_{\rm J/\psi}}{\sqrt{\slfrac{\left(1+x_2\right)}{x_2}}}$).

\begin{figure}[htbp]
\centering
\resizebox{0.8\textwidth}{!}
{\includegraphics*[bb=0 0 539 383]{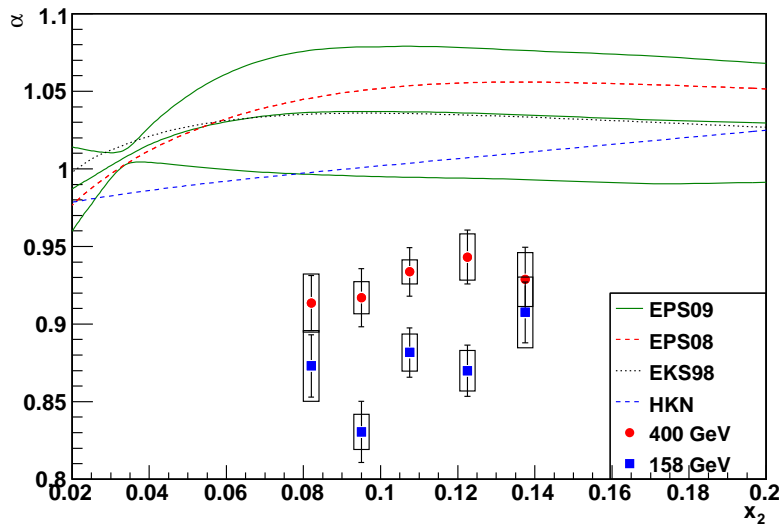}}
\caption{$\alpha$ as a function of $x_2$, for the 158 and 400 GeV data samples. Also shown in the plot is a calculation of $\alpha$
as resulting from shadowing alone. Various shadowing parameterizations have been considered.}
\label{fig:shad}
\end{figure}

Since our results at the two energies cover, with good 
approximation, the same $x_2$ domain, we have performed an analysis of nuclear effects in five $x_2$ bins, covering the region 
0.08 $<x_2<$ 0.14.
In Fig.~\ref{fig:shad} we plot $\alpha$ as a function of $x_2$ for the two energies under study. We also show in the same plot the
expected $\alpha$ value that would be obtained if shadowing were the only nuclear effect to be present. Various parameterizations of
nuclear shadowing have been considered~\cite{Esk09,Hir04,Esk99,Esk08}, including the recent EPS09 result, where an estimate of the uncertainty
of the calculation was also carried out. 
Looking at Fig.~\ref{fig:shad} we see that shadowing alone would lead to $\alpha$ values larger than 1 in our $x_2$ acceptance. 
This result implies that the other nuclear effects produce a significant suppression of the J/$\psi$ yield.

\begin{figure}[htbp]
\centering
\resizebox{0.8\textwidth}{!}
{\includegraphics*[bb=0 0 539 383]{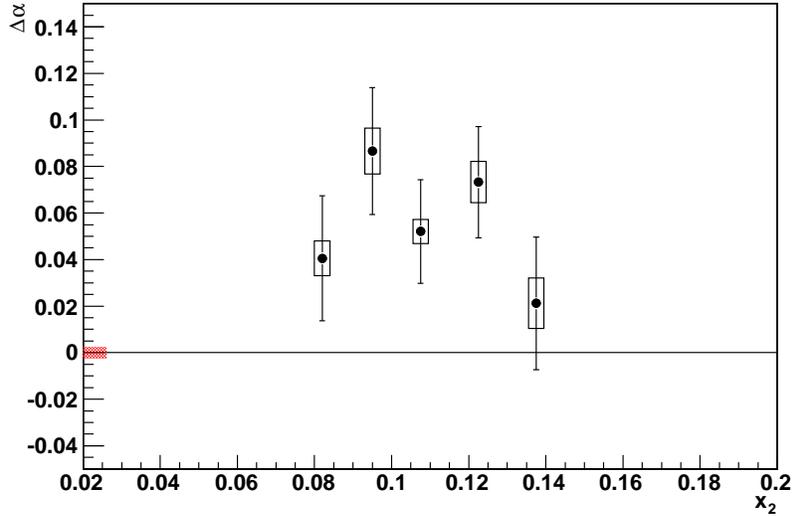}}
\caption{$\Delta\alpha$=$\alpha(\rm 400 GeV)-\alpha(\rm 158 GeV)$, as a function of $x_2$. The empty boxes around the points represent the
uncorrelated systematic uncertainty. The systematic error common to all the points is shown as a filled box around the $\Delta\alpha=0$ line.}
\label{fig:4}
\end{figure}

The other remarkable feature emerging from Fig.~\ref{fig:shad} is that $\alpha$ is not the same at a fixed $x_2$, for the two values
of $\sqrt{s}$. In our analysis, the systematic uncertainties related to the measurement of $\alpha$ are partly correlated, in
particular those corresponding to the measurement of the J/$\psi$ detection efficiency. Therefore, in Fig.~\ref{fig:4} we plot, as 
a function of $x_2$, the quantity $\Delta\alpha=\alpha(\rm 400 GeV)-\alpha(\rm 158 GeV)$ which is affected by a significantly
smaller systematic uncertainty. 
The results clearly indicate that $\Delta\alpha\neq 0$ (in the covered $x_2$ range we have
$\langle\Delta\alpha\rangle=0.056\pm0.011(\rm{stat})\pm0.004(\rm{syst})$), and that $\Delta\alpha$ does not vary appreciably in the $x_2$ 
region under study.

Having factorized the effect of shadowing, which is the same at the two energies at fixed $x_2$, the observation of $\Delta\alpha>0$ could be attributed to a stronger final state interaction at 158 GeV. Since, for a given $x_2$, the kinematics of the collision of the \jpsi\ with the target nucleons is the same at 158 and 400 GeV, the observed effect may be related to a change in the \jpsi\ break-up cross section at fixed $\sqrt{s_{\rm J/\psi N}}$.
A different weight of the color-octet and color-singlet precursor $c\bar{c}$ states in the production process between the two energies could be a natural explanation.
However, Non-Relativistic QCD (NRQCD) calculations carried out in the fixed target energy domain~\cite{Vogt00}, only show a weak $\sqrt{s}$ dependence of the color-octet and color-singlet relative contributions.
Therefore, a strong variation of the \jpsi\ break-up cross section with $\sqrt{s}$ seems unlikely.
In such a case one should observe, for the two incident proton energies, the same $\alpha$ at a given $x_2$ in contrast with our observation.
In this scenario, processes different from final state break-up and parton shadowing must be advocated in order to explain the observed nuclear effects on \jpsi\ production. In particular, effects such as the initial state energy loss of the incident parton might play an important role and deserve further investigation.  
 
\section{Conclusions}
\label{sec:concl}

The NA60 experiment has measured \jpsi\ production in \mbox{p-A} collisions at the CERN SPS, at 400 and 158 GeV, 
the latter energy being the lowest one where a detailed systematic study has been performed. 
The results show a suppression of the \jpsi\ yield in cold nuclear matter, which is larger at lower incident energy. 
A comparison with results from previous experiments indicates that nuclear effects on \jpsi\ production, at constant $x_{\rm F}$, 
exhibit a strong $\sqrt{s}$-dependence. 
The observation of different $\alpha$ values at the two energies, for a constant $x_2$, may suggest a strong change in the \jpsi\ break-up cross section and/or the presence of other effects like initial state energy loss.
These data can also provide an important baseline for heavy-ion collision studies.








\end{document}